\documentclass[twocolumn,astrosymb]{aastex631}
\usepackage{mathptmx}
\usepackage{CJK}

\newcommand{\citeeg}[1]{\citep[e.g.,][]{#1}}
\graphicspath{{./}{figures/}}

\begin{document}

\title{OSSOS. XXIX. The Population and Perihelion Distribution of the Detached Kuiper Belt}

\begin{CJK*}{UTF8}{gbsn}
\author[0000-0001-6597-295X]{Matthew Beaudoin}
\author[0000-0002-0283-2260]{Brett Gladman}
\author[0000-0003-1215-4130]{Yukun Huang (黄宇坤)}
\affiliation{Department of Physics \& Astronomy, University of British 
Columbia, 6224 Agricultural Road, Vancouver, BC V6T~1Z1, Canada}
\author[0000-0003-3257-4490]{Michele Bannister}
\affiliation{School of Physical and Chemical Sciences --- Te Kura Mat\=u
University of Canterbury, Private Bag 4800, Christchurch 8140, New Zealand}
\author[0000-0001-7032-5255]{J.J. Kavelaars}
\affiliation{Herzberg Astronomy and Astrophysics Research Centre, 
National Research Council, 5071 West Saanich Road, Victoria, BC V9E~2E7,
Canada}
\author[0000-0003-0407-2266]{Jean-Marc Petit}
\affiliation{Institut UTINAM UMR6213, CNRS, Universit\'e Bourgogne Franche-Comt\'e, OSU Theta F-25000 Besa\c con, France}
\author[0000-0001-8736-236X]{Kathryn Volk}
\affiliation{Lunar and Planetary Laboratory, The University of Arizona, 
1629 E. University Blvd., Tucson, AZ 85721}



\begin{abstract}
The detached transneptunian objects (TNOs) are those with semimajor axes beyond
the 2:1 resonance with Neptune, which are neither resonant nor scattering.
Using the detached sample from the OSSOS telescopic survey, we produce the first studies
of their orbital distribution based on matching the orbits and numbers of the
known TNOs after accounting for survey biases.
We show that the detached TNO perihelion ($q$) distribution cannot be uniform, but
is instead better matched by two uniform components with a break near $q\approx40$~au.
We produce parametric two-component models that are not rejectable by the 
OSSOS data set, and estimate that there are 
36,000$^{+12,000}_{-9,000}$ detached TNOs with absolute magnitudes $H_r < 8.66$
($D \gtrsim 100$~km) and semimajor axes $48 < a < 250$~au (95\% confidence 
limits).
Although we believe these heuristic two-parameter models yield a correct population
estimate, we then use the same methods to show that the perihelion distribution 
of a detached disk created by a simulated rogue planet matches the $q$ distribution
even better, suggesting that the temporary presence of other planets in the 
early Solar System is a promising model to create 
today's large semimajor axis TNO population. 
This numerical model results in a detached TNO population estimate
of 48,000$^{+15,000}_{-12,000}$.
Because this illustrates how  difficult-to-detect 
$q>50$~au objects are likely present,
we conclude that there are 
\mbox{$(5 \pm 2)\times10^4$} dynamically detached TNOs,
which are thus roughly twice as numerous as the entire transneptunian
hot main belt.
\end{abstract}


\section{Introduction}\label{sec:intro}
\end{CJK*}
Transneptunian objects (TNOs) are considered to be leftovers from the early stages of
planet formation in the Solar System, when the Sun was still surrounded by a
protoplanetary disk.
By the time the four giant planets formed and then the 
young Sun's solar wind had expelled the gas and remaining dust, 
planetesimals existed all across the outer Solar System.
At the very least, this included millions of objects with 
diameters $D>50$~km, all the way up to 
a set of dwarf-planet sized objects (reaching a few thousand 
kilometres across, such as Pluto), and then a uniform number of planetary objects.

The most-discussed paradigm is that the TNO region beyond Neptune
consists of two components. The ``cold classical Kuiper belt'' has various historical
definitions, but is recently commonly restricted to the main belt's low inclination 
component, which seems to only exist on low-eccentricity orbits from semimajor axes 
$a=42.4$--47.5 au.
This current cold belt has been suggested  \citep{Kavelaars2021}
to be a largely unaltered remnant preserving the original 
formation size distribution and cold TNO number 
(of approximately $10^4$ objects with $D>100$~km).
In contrast, the more numerous ``hot'' population
(hot in terms of orbital eccentricity $e$ and inclination $i$) is generally
thought to have formed closer to the Sun and then scattered
out with some small fraction decoupled from Neptune's 
influence and preserved to the present day 
\citep[see reviews by][]{Morbidelli2008, Nesvorny2018, Gladman2021}. 
In this scenario, all the hot populations from Neptune trojans
(co-orbitals librating around Lagrange points)
to the Oort cloud share a common origin.
\citet{Petit2011} showed that the number density of objects
from 30 to 100 au across the inner belt (closer to Neptune
than the 3:2 resonance), main belt (from the 3:2 to 2:1) and the 
region beyond the 2:1 at 48~au could be smoothly connected.
Beyond 48~au there is either no cold belt or there is a sudden dramatic drop in the
surface density \citep{Gladman2021};
the stable TNOs consist of only those hot-population 
objects trapped in distant mean-motion resonances (MMRs) and the 
``detached'' population.
After initial scattering to large $a$, the detached 
objects had their perihelia distance $q$ raised by unclear 
processes to values beyond which Neptune can strongly influence 
them today. 

In this paper, we define ``detached'' objects as the TNOs beyond the Neptune 2:1 mean-motion
resonance ($a > 48$~au) that are neither (mean-motion) resonant nor scattering as per the 
classifications described in \citet{Gladman2008nomen}.
Scattering objects have semi-major axes that can be significantly altered 
by gravitational interactions with Neptune on Myr timescales \citep[see][]{Gladman2008nomen}.
Roughly speaking, detached TNOs have large enough perihelia
to be dynamically decoupled from Neptune, thus avoiding significant gravitational interaction
(scattering) on 10 Myr to Gyr timescales; the amount of mobility and the scattering time scale depend on $a$ and $q$
\citep[see][]{Gladman2002, Bannister2017, Khain2020, Batygin2021}. 
Other dynamical classes of TNOs have origins and emplacement mechanisms that are mostly understood 
and only require the action of the known planets \citeeg{Duncan1997, Morbidelli2008, Malhotra2019}.
The physics of perihelion lifting for detached TNOs are unclear, but there are many ideas,
including dynamical diffusion \citep{Bannister2018},
dropouts from  mean-motion resonances during grainy Neptune migration  
\citep{Kaib2016, Nesvorny2016} and/or Neptune's orbital circularization phase \citep{Pike2017a}, 
interactions with a distant giant planet \citep{Gomes2006, Lawler2017},
a stellar flyby \citep{Morbidelli2004},
perturbations in the Solar birth cluster \citep{Brasser2015},
and the action of a rogue planet that was scattered to the outer Solar System and
lifted perihelia before being ejected \citep{Gladman2002, Gladman2006, Silsbee2018}. 

In the way this paper (and past papers from the CFEPS/OSSOS collaboration 
which estimate populations) approach the problem, if one is going to compare a
cosmogonic numerical simulation to observations, then one determines the 10 
Myr behaviour of the simulation's orbits at the current epoch, and divides 
them exactly the same way as the dynamical classification of observed objects.  
That is, one determines which of the simulation's particles are resonant and 
which are scattering, on a 10 Myr interval, and the remainder are detached.
Thus when we provide a measure of a ``detached population'', we are providing 
the population of non-resonant TNOs beyond the 2:1 that are not scattering 
on 10 Myr time scales.

\begin{figure*}
    \centering
    \includegraphics[width=0.8\textwidth]{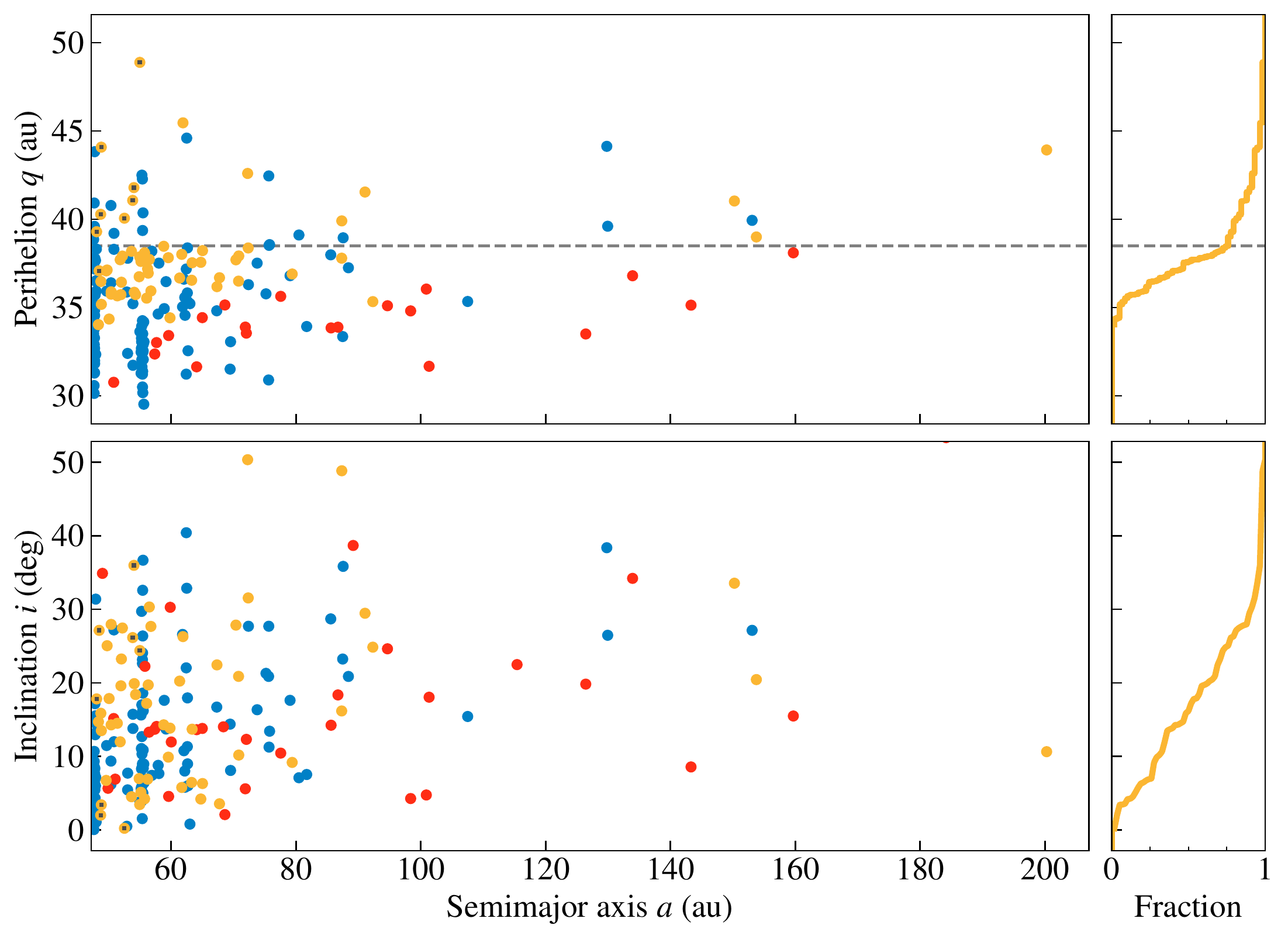}
    \caption{The detached (gold points) TNO sample from the OSSOS survey.  
    The \mbox{$q=38.5$~au} horizontal dashed line in the upper $a,q$ plot highlights a break
    in cumulative perihelion distribution (upper right plot).
    The eight OSSOS ``outer classicals'' we adopt as detached (see text) are indicated 
    by overlaid black squares.
    Scattering objects (red) and objects in mean-motion resonances (blue) are diagnosed via 10~Myr numerical integration and are plotted for reference.
    Three obvious resonant semimajor axes are 48~au (2:1), 55~au (5:2) and 63~au (3:1); the resonances stabilize even $q<34$~au TNOs.
    Note there are strong detection biases against objects in the upper right of each of the two distributions.
    The cumulative distributions (right panels) are for only the detached sample.
    The inclination distribution (bottom) is not obviously different among the three populations, although the 2:1 resonance
    is relatively poor in large-$i$ TNOs \citep{Chen2019}.}
    \label{fig:aqdynclass}
\end{figure*}

There is only one observational estimate of the number of dynamically detached objects 
in the literature.
As one result of the Canada-France Ecliptic Plane Survey (CFEPS), \citet{Petit2011}
estimated the number
of detached objects with $a > 48$~au to be $N(H_g \le 8) = \text{10,000}^{+7,000}_{-5,000}$, with the uncertainties representing a 95\% confidence interval. 
It is important to note that
\citeauthor{Petit2011} used a very small sample of only 13 detached objects.
Due to this paucity of objects, they adopted to use the same
$q$ and $i$ distributions as for the hot main belt, which
are purely empirical models based on the observed distribution 
of hot main-belt classical TNOs. 
Thus there is currently no population estimate of the detached
objects with independently-determined orbital element distributions. 
Based on the non-rejectable parametric models we describe in this article,
we produce such an estimate. 
Population estimates serve as a goalpost for emplacement models, in the sense that any model 
must reproduce the intrinsic population to be considered valid. 
The relative population of associated dynamical classes
are also of interest for physical models; for example: what
is the ratio between the detached and resonant populations in 
the $a$ = 48--250~au range?

In this article, we make use of the TNO catalogue from the OSSOS++ collections of characterized surveys, comprised of the Outer Solar System Origins Survey \citep[OSSOS;][]{Bannister2016, Bannister2018}, the Canada-France Ecliptic Plane Survey \citep[CFEPS;][]{Petit2011}, the CFEPS High-latitude Component \citep[HiLat;][]{Petit2017}, and the \citet{Alexandersen2016} survey.
With a vastly increased sample of detections resulting from the OSSOS++ suite, we can now quantitatively explore the orbital element and size distributions of the detached objects from their inner boundary at the Neptune 2:1 mean-motion resonance ($a\simeq48$~au) to $a = 250$~au. 
The OSSOS collection provides 58 objects that numerically demonstrate detached behavior (non-resonant and non-scattering on 10 Myr time scales)
having perihelia $q$ = 34.03--48.89~au; no $q < 34$~au OSSOS object
demonstrates detached behavior.
For the purposes of this manuscript we included, in our set of $a > 48$~au
detached objects, eight ``outer classicals'' (see \autoref{fig:aqdynclass}) 
having $a$ = 48--55~au with eccentricity $e < 0.24$ \citep[see][]{Bannister2018}. 
According to the \citet{Gladman2008nomen} classification scheme, non-scattering 
and non-resonant objects beyond the 2:1 are nominally detached only if they have
$e > 0.24$; for this work, we are adopting the viewpoint that these 
eight OSSOS outer classicals were emplaced by the same mechanisms as the other
detached TNOs, so we include them in our list.
Motivated by the largest $a$ object in the OSSOS++ detached sample, 2013~UT$_{15}$ (OSSOS designation \texttt{o3l83})  with  $a \simeq 200$~au, and the relative sparsity of larger-$a$ objects, we will thus only give orbital element, $H$ magnitude, and population constraints for objects with $a < 250$~au.
Beyond this there are very few objects, especially at large $q$, which were found in a large variety of different Solar System surveys which we are unable to rigorously debias.
This includes the realm of the so-called ``extreme TNOs'' 
(with only vaguely-defined motivations for the inner $a$ boundary);
that population is in addition to what we measure in this manuscript,
and could include a significant population of
$q\gg50$~au TNOs (like Sedna and 2012~VP113, with 
$q\simeq75$ and 80~au respectively) but with $a > 250$~au.

\autoref{fig:aqdynclass} shows the $(a, q)$ and $(a, i)$ distributions of TNOs in this semimajor axis range. 
Histograms at the right of each panel show the cumulative distributions in $q$ and $i$;
one obvious feature in the $q$ distribution of the detached objects is the apparent rollover at $q \simeq 38.5$~au. 
Can this feature be explained by observational bias and preferential detection of low-$q$ objects? 
Or rather is it real and some kind of indicator for the perihelion lifting mechanism? 
To answer these questions, we use the OSSOS Survey Simulator \citep{Petit2011, Petit2018} which applies the observational bias, from which the real TNO detections suffer, to a given model \citep{Lawler2018ss}. By comparing the simulated detections to the real objects, we are thus able to evaluate the suitability of various models for the detached objects' intrinsic $q$ distribution with a focus on the aforementioned rollover.

We first explore empirical models that are non-rejectable matches to the distribution of observed detached TNOs, providing an estimate of the
intrinsic detached population.
We then briefly study an example of matching a cosmogonic numerical simulation of the orbital distribution that results from the hypothetical
presence of an additional rogue planet 
(one formed in Solar System that is eventually ejected via gravitational scattering),
showing that the resulting
observationally-biased perihelion distribution from that model
bears a striking resemblance to the known detections.

\medskip

\section{Survey Simulations of Orbit and Size Distributions}\label{sec:surveysim}
The concept of a survey simulator is straightforward: given intrinsic
distributions of orbital elements, the software creates an object by randomly
drawing from them, places it on the  sky, applies all of the
observational biases of a characterized survey, and reports if the object
would have been detected \citep[for in-depth discussions, see][]{Kavelaars2009,Lawler2018ss}. Providing the survey simulator
with parametric distributions allows the model to be attenuated by the observational biases.
To do this, 
a synthetic object that is randomly drawn from the intrinsic distribution is exposed
to the biases of the survey, including field-of-view (it must be within the
survey's coverage), CCD filling factor, rate cuts, tracking fraction (near
unity for OSSOS, the largest survey in our ensemble), and
detection efficiency. For example, a TNO may be drawn at a sky position
that happens to be in a field of the survey, but its magnitude may be too faint
for detection. The most important criterion for detection, after the on-sky location, is the
object's magnitude; whether or not the object is detected in a field depends
heavily on the detection efficiency function of the observation block/field.

First, we provided parametric distributions of $a$, $q$, $i$, and $H$ to the survey simulator, and the output is a list of properties of the
objects that were drawn and those that were tracked.
The collection of tracked objects comprise our ``simulated detections''. 
The survey simulator is usually set to run until a large number of simulated objects 
are found to ensure well-sampled cumulative distribution functions (CDFs) to which we 
compare the small set of real detections.

Because these models include difficult-to-detect TNOs, 
the simulator must draw enormous numbers of randomized intrinsic objects 
(most of which are then not observed) to get
a statistically useful number of simulated detections.
In our models, millions of candidate
TNOs must be examined before $\sim$5,000 tracked detections are made. 
This ratio is indicative of how much larger the true population is than 
the detected number.

The survey simulator can also operate in ``lookup'' mode (as
opposed to ``parametric'' mode). In this case, instead of randomly drawing from
distributions, it draws objects from a lookup table and exposes them to the
survey biases; such lookup tables are most often the result of an orbital simulation with physical motivation. 
The OSSOS Survey Simulator can run until
it reaches the end of the lookup table, or until a specified number of simulated
detections are obtained. This can lead to drawing the same model TNO multiple times;
this is often dealt with by randomizing angular variables and by implementing a
very small ``fuzzing'' of orbital elements (no more than $\pm0.5\%$) to allow
for more simulated detections than a model might provide.
In \autoref{sec:rogue}, we forward-bias the results of a rogue planet
scenario integration using the survey simulator in lookup mode.

Following methodically similar projects in the literature
\citep{Kavelaars2009,Shankman2013,Shankman2016,Alexandersen2016,
    Bannister2016,Lawler2018}, we examine the rejectability of a model
using a ``bootstrapped'' Anderson--Darling (AD) 
goodness-of-fit
test \citep{Anderson1954}, which tests the hypothesis that 
a sample originated from a specified population. 
The AD statistic is conceptually similar to the Kolmogorov--Smirnov (KS)
statistic (which is the extremum of the difference
between the two distributions).
We prefer the AD methodology for our tests as it is more sensitive to 
differences at the tails of the distributions than the KS test.
In what follows, in all but one case we use the AD test to
establish confidence intervals around the model parameters, and
``reject'' a parameter set when the AD probability is $p<0.05$.

In order to estimate model rejectability, one must ``bootstrap'' the distribution
of AD statistics \citeeg{Kavelaars2009, Gladman2012, Shankman2016, Lawler2018}.
We run the survey simulator to generate a set of simulated detections (typically 3,000--10,000) much larger 
than the number of known objects.
First, the ``real'' AD statistic between the observed objects and simulated
detections from a model is calculated. 
We then select a subsample randomly from the simulated detections
that is the same size as the number of real objects, computing the AD statistic
between it and the simulated distribution. By repeating this process many
thousands of times, one bins them to estimate the bootstrapped AD
statistic distribution. We then look
at the probability of obtaining an AD statistic greater
than the ``real'' AD statistic: if $<$5\% of the simulated statistics are
greater than the real statistic, and thus the probability is $<$5\%, we
conclude that the model distribution is
inconsistent with observations and reject it at the 95\% confidence level
(or similarly, $<$1\% determines 99\% confidence).
We refer to this probability as the ``AD test
result/probability'' or just as probability $P$. This quantity can be thought 
of as an ``acceptability criterion'', or $1-P$ as the rejection confidence.

The above process applies to each
parameter of interest ($a$, $q$, $H$, $i$); here we treat each parameter as
independently rejectable. It is possible to run a multi-parameter AD
model rejection test \citeeg{Alexandersen2016, Lawler2018ss}, but
correlations between parameters (especially $a$, $q$, and $H$) make it difficult
to find specific areas of issue that cause the entire model group to be rejected.
This is more applicable to parametric models where each
distribution is tweaked to fit a known population. When forward-biasing the
results of a numerical integration, it can be useful
to have a total rejection criterion. However, we maintain single-parameter
AD testing throughout to attempt to identify specific discrepancies
in particular orbital elements.

When considering implantation into the Kuiper belt, we find that the 
$q$ distribution is the critical element, as it measures how much
perturbation is needed to convert a Neptune-coupled orbit to one that is 
detached (since TNOs are scattered to a very large $e$ range but a much
more limited $q$ range).
Thus the $a$ and $e$ distributions are largely set by basic Neptune scattering,
while the $q$ distribution is much more sensitive to the detachment physics.
The $e$ distribution of the scattering and detached TNOs is so broad it 
mutes the signature of the critical perihelion distribution, and because 
the $q$ range is much narrower than the $a$ and $e$ range, the latter two
quantities are very correlated and so testing each provides nearly the same 
statistical information. For this reason, we examine the $q$ distribution 
instead of the $e$ distribution and find that this choice allows for tighter 
constraints on non-rejectable orbital element distributions.

Since the overall goal of the survey simulator is to compare model orbital and
size distributions of the outer Solar System to distributions as
observed by a survey, we must develop intrinsic distributions to test. We will test a number of
potential detached object distributions for both size ($H$ magnitude) and
orbital elements guided by previous studies
\citep{Shankman2013, Shankman2016, Lawler2018}. We first find $i$, $a$, and $H$
distributions for the detached objects that are non-rejectable, basing them on
models describing other TNO populations. Since we do not know the
dynamics behind the detached objects' perihelion lifting, there is no
expectation for
the shape
of the intrinsic $q$ distribution. We gradually
increase the complexity of models up from the simplest case, a uniform $q$
distribution, looking for models that we cannot reject by the statistical
method described above. Then, in \autoref{sec:popest}, we use a
non-rejectable parametric model to estimate the population of the detached
objects.

\subsection{Inclination Distributions}\label{sec:idist}
\begin{figure}
    \center
    \includegraphics[width=\columnwidth]{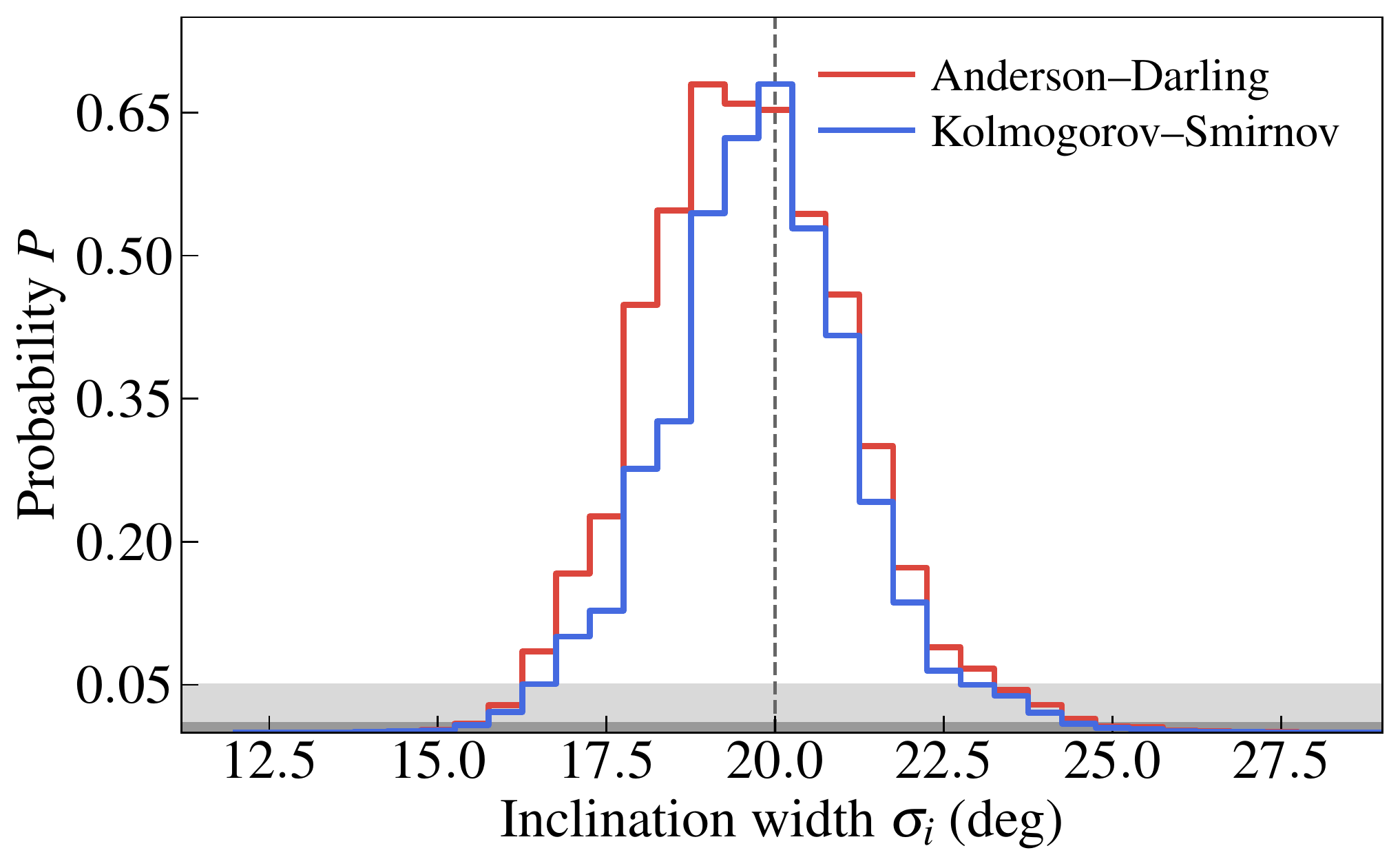}
    \caption{Bootstrapped probability of the width $\sigma_i$ of an intrinsic detached-TNO inclination distribution of the form $\sin i \times \exp \left[ -i^2/(2\sigma_i^2) \right]$.
    Light and dark grey horizontal bands at bottom denote rejection at 95\% and
    99\% confidence intervals, respectively.
    Both the KS and AD tests for inclination agree that a model with a width 
    $\sigma_i\simeq 20^\circ$ is a good representation of the 
    intrinsic distribution and reproduces the OSSOS detections.
    The tests shown were performed with the least-rejectable two-component $q$ distribution
    as described in \autoref{sec:splitq}.}
    \label{fig:iwidth}
\end{figure}

The main-belt TNO inclination distribution is well described
by $\sin(i)$ multiplied by overlapping Gaussians \citep{Brown2001}. 
Individual populations outside the main belt are often fit by 
$\sin(i)$ times a single Gaussian,\footnote{Functional forms other than
Gaussians are described by \citet{Gulbis2010}.} with a width 
$\sigma_i$ \citeeg{Petit2011, Petit2017, Gladman2012, Bannister2018},
of the form
\[ 
P(i) \propto \sin(i) \exp\left(\frac{-i^2}{2\sigma_i^2}\right).
\]

In our parametric models, we find that permitted values of the $i$ distribution width 
are largely independent of the assumed $a$ and $q$ distributions. 
Testing models with very different $\sigma_i$ only modifies the AD test result of the $a$ 
and $q$ distributions by a few percent at most compared to the least-rejectable 
$\sigma_i$, and in no cases does it shift an acceptable result to a rejectable one.
With this established, we want to determine a nominal $\sigma_i$ to use in
our models and treat it as a fixed value when investigating the $q$ distribution.
To do so, we run the survey simulator for $\sigma_i=12^\circ$--$31.5^\circ$ 
with a step size of $0.5^\circ$. 
For each iteration, we generated 10,000 simulated detections and computed the
AD test result.
In \autoref{fig:iwidth}, the $y$ axis is 
the probability $P$ that a bootstrapped AD statistic for a model with width 
$\sigma_i$ is greater than the real AD statistic. 
We reject models with $P < 5\%$ and $P < 1\%$ with
95\% and 99\% confidence, respectively.
We find a least-rejectable inclination width
of $\sigma_i = (20 \pm 3)^\circ$, where the uncertainties cover the 95\% confidence interval.

For comparison, the hot classical $i$ distribution can be described by the same 
function with $\sigma_i = 14.5^\circ$ \citep{Petit2017}. 
Likewise, scattering TNOs have been modelled with an initial $\sigma_i = 
12^\circ$, which is then dynamically eroded, increasing the width
by a few degrees \citep{Shankman2013, Shankman2016}.
Our result of $\sigma_i = 20^\circ$ suggests a somewhat hotter inclination distribution for detached objects than for the hot main belt TNOs and the scattering TNOs, but is comparable to distant TNOs in large-$a$ resonances 
\citep{Crompvoets2022}. 
In particular, a wider $i$ distribution than the hot classicals implies that
the detached objects become more dispersed in inclination than the population 
that was initially scattered out. That is, the $q$-raising process that 
detaches TNOs likely simultaneously raises their inclinations \citeeg{Gomes2008}.

\subsection{Semimajor Axis and Absolute Magnitude Distributions}
\label{sec:aandHdists}
For the models in this paper, the semimajor axis is differentially 
distributed according to a power law proportional to $a^{-\beta}$, where we take
$\beta = 2.5$. This index for the $a$ distribution is characteristic
of an early scattering disk \citep{Huang2022}.
As further motivation for this choice, \citet{Petit2011} find that a single, 
continuous primordial hot population distributed according to the $a^{-2.5}$
power law can account for the inner belt, the hot main belt, and the detached TNOs.

The lower $a$ limit for our study is based on our definition of detached objects. 
For that population, the lower limit on $a$ can be taken to be the outer 
edge of the 2:1 MMR at $a_\mathrm{min} = 48$~au \citep{Gladman2008nomen}.
Based on the largest-$a$ detached TNO in the sample and to insure the large-$a$
tail of the simulated detections is statistically robust,
we adopted an upper a limit $a_\mathrm{max} = 250$~au.
There are only a few (3--5) detached objects with $a > 250$~au, and much further than that passing stars and galactic tides are non-trivial at the object's aphelion
\citep{Sheppard2019}, which may alter the power law. 
If the hypotheses that a massive planet still exists in the outer Solar System are
true \citep{Trujillo2014, Batygin2019}, the semimajor axis distribution will be modified
\citep{Lawler2017}.
Given our interest in the apparent perihelion break 
around $q \simeq 38.5$~au (which is only sampled by OSSOS at low $a$), 
the 250~au upper limit on $a$ is appropriate.
Due to no OSSOS detections, our study is insensitive to the orbital distribution 
beyond that in any case.

We found that this semimajor axis distribution is rarely rejectable
for any of our chosen $H$, $i$, and $q$ distributions. 
This may be surprising because it implies that the early scattered
disk (from which the detached objects being lifted) may not have 
reached the expected long-term steady state of $dn/da \propto a^{-1.5}$.
We will return to this topic in \autoref{sec:discussion}.

The absolute magnitude ($H$) distribution is motivated by studies of other TNO
populations. Throughout this manuscript $H$ magnitudes are reported in the $r$ band.
In general, simple $H$ distributions often take the form of a
base-10 exponential law with a cumulative distribution function of the form
$N(<\!H) \propto 10^{\alpha H}$, where $N(<\!H)$ is the number of objects
with absolute magnitude less than $H$ and $\alpha$ is the ``slope'', referring
to the distribution's appearance on a logarithmic scale.\footnote{A cumulative 
diameter distribution $N(>\!D)\propto D^{-Q}$ corresponds to an $H$ distribution 
with $\alpha=5Q$.}
It is common that a single slope $H$ distribution is inadequate for describing
a TNO population. 
Models with different slopes over different $H$ ranges include
broken or rolling power laws that transition to shallow slopes at large $H > 8.5$
\citeeg{Bernstein2004,Fraser2014,Lawler2018}.
However, at the bright end it has long been apparent \citeeg{Brown2008}
that the dwarf planets are over abundant at $H<3$, so the ``break'' is in
the opposite sense.
Here we adopt a broken power law with two independent slopes: a shallow
$\alpha_{dp}$ for the bright (dwarf planet) end of
the distribution, and a steeper $\alpha_*$ for $H > 3$, separated 
by some transition magnitude $H_\mathrm{break}$. 
\citet{Ashton2021} found that all known large TNOs could be represented by a two-component
exponential law with $\alpha_{dp} = 0.14$, $\alpha_* \simeq 0.6$, with a break
at $H_\mathrm{break} = 3$. We adopt these parameters for the detached $H$ 
distribution, since there is no detached-specific distribution in the literature
and we have almost no sensitivity to $H > 8.5$.

Although we used a minimum value of $H=0$, we show below that 
OSSOS is insensitive to the detached dwarf planet regime%
\footnote{Eris is a prominent example.} since not enough
detached TNOs were detected; they make up $\ll1$\% of a detected
sample. 
Since so few bright objects are drawn, it does not make a 
difference if the bright limit is set to $H = 0$ or down to 
Pluto's magnitude $H \simeq -3$, for example. 
We set a faint-end $H$ limit  of
$H = 9$, which is $\sim$0.5 magnitudes fainter than the point at which one
can no longer trust the survey simulator's assessment of 
detection efficiency.
Objects drawn beyond $H = 8.5$ are, on average, at the limit of survey
sensitivity, and the sharp, exponential drop-off in detection efficiency makes
the survey simulator sensitive to small variation in $H$. 
The faint-end limit is set fainter than the limit we will use for our population 
estimate ($H_r = 8.66$), since the observed $H$ of a simulated object can be
brighter or fainter than the intrinsic $H$ as drawn from the distribution due to
the simulator modelling the photometric variation of the 
modelled objects.

If the object is marked
as ``detected'', the apparent magnitude at detection is converted to a surmised 
absolute magnitude $H_\mathrm{sur}$ using the
heliocentric distance. 
We do not know the intrinsic $H$ magnitude for any real detached TNO.
Rather, observations can only tell us the surmised magnitude. Similarly, simulated detections 
can appear brighter or fainter than they ``really'' are, and our simulated 
detections at the faint limit rely on the correctness of the survey simulator's 
low signal-to-noise photometric scatter models.
We found that the $\alpha_* = 0.6$ intrinsic distribution does an admirable job of representing the $H$ magnitudes of the detected
detached TNOs for our models down to the detection limit near $H \simeq 8.5$.

\section{Uniform Perihelion Distribution}\label{sec:unifq}
Recall the question of the ``perihelion break'' as introduced in
\autoref{sec:intro} and illustrated in \autoref{fig:aqdynclass}. In the
cumulative distribution, this appears as a sharp change in the slope around
$q\simeq38.5$~au. 
We wish to create a parametric model describing the detached
object's $q$ distribution to forward bias using the survey
simulator, testing whether this break could be entirely due to bias 
or if it is instead a feature in the intrinsic population.

\begin{figure*}
    \centering
    \includegraphics[width=0.85\textwidth]{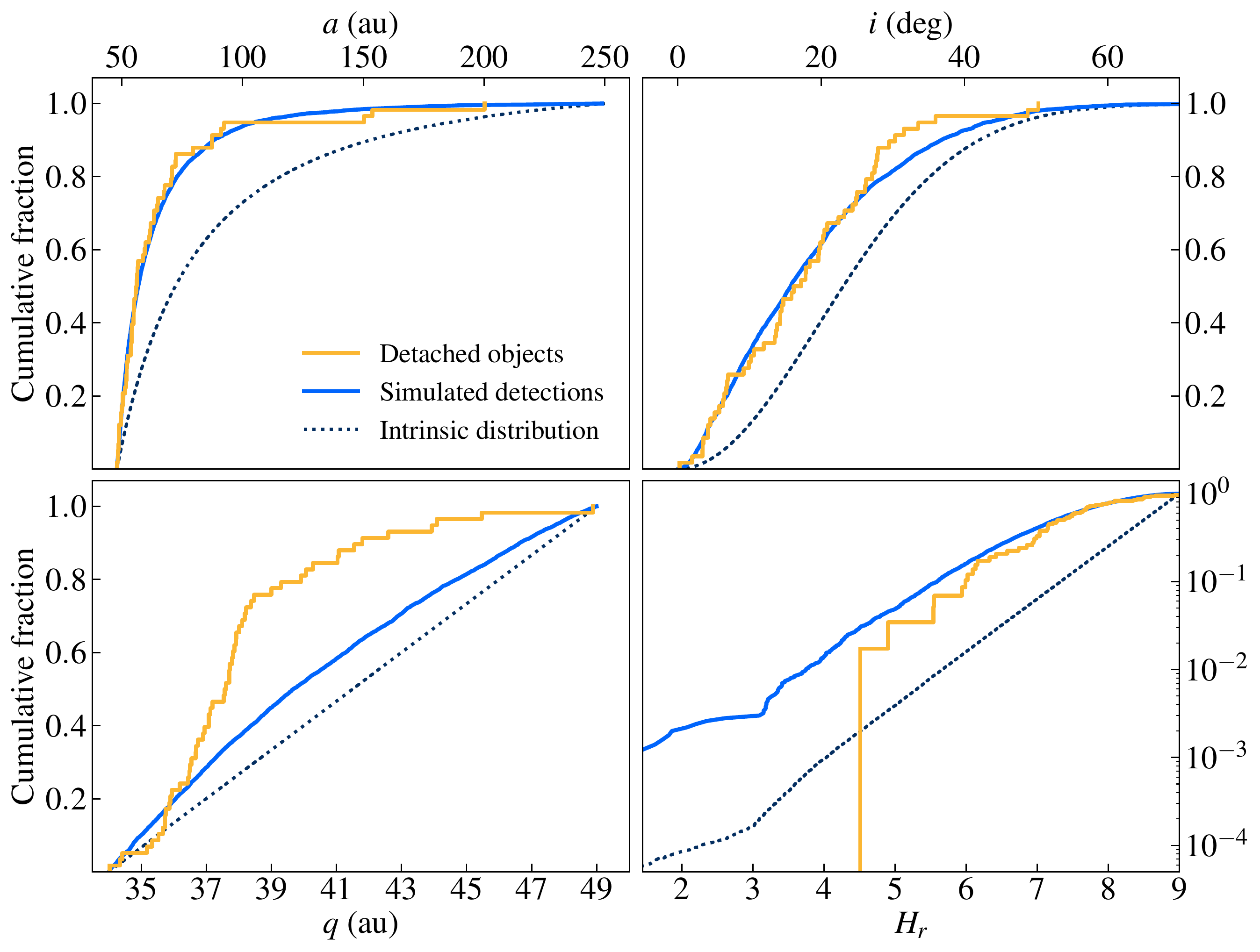}
    \caption{Cumulative distributions for a straw man model with a uniform perihelion distribution.
    Each cumulative panel shows the OSSOS detached sample (gold), the intrinsic
    model (dotted black curves) and the simulated detection distribution from that intrinsic model after biasing by the OSSOS survey simulator.
    Although the $a, q,$ and $H_r$ distributions are not 
    statistically rejectable,
    the lower left panel shows that despite a mild preference for detecting 
    low-$q$ detached objects, the detection biases cannot produce such a dramatic
    concentration to $q<38.5$~au (with an AD bootstrapped probability of drawing the observed sample from this model of $<$0.01\%).}
    \label{fig:unifqcumu}
\end{figure*}

We first test the case of a uniform distribution in perihelion,
expecting to strongly reject it. One does expect the observational
bias to favor low-$q$ detections and shift the constant-slope CDF
in $q$, but whether it can alter the distribution to the extent
observed in the real objects requires survey simulation.
The smallest and largest $q$ detached TNOs in our sample have $q
= 34.03$~au and $q = 48.89$~au. To cover the full range of detached 
objects, we thus fix $q_\mathrm{min} = 34$~au and $q_\mathrm{max} = 49$~au. 
The lower limit is roughly the same as the minimum $q$ 
a detached object could have without scattering.
In these parametric models,
we are only attempting to model the perihelion distribution at 
a heuristic level;
we primarily desire a non-rejectable model to produce a 
population estimate
(and we show below that this population estimate is nearly independent of the details of the assumed $q$ distribution).

\autoref{fig:unifqcumu} shows the cumulative distributions of the real 
detached objects and the survey simulator output for the case with intrinsic $q$ uniformly distributed between 34~au and 49~au.
This uses the nominal parametric models for $i$, $a$, and $H$ as described 
in \autoref{sec:idist} and \autoref{sec:aandHdists}.
The semimajor axis and inclination distributions show the expected strong detection biases
to smaller values; without a survey simulator the magnitude of this bias is difficult
to estimate.
The intrinsic $a^{-2.5}$ power law and the Gaussian inclination width of $\simeq$20$^\circ$
match the OSSOS detections well, with AD probabilities of 91\% and 62\%, respectively.
The $H_r$ magnitude distribution AD probability is 57\% which is also completely acceptable
although the eye can see that $H_r<7$ detections are slightly overproduced with this
orbital and $H_r$ magnitude distribution.
The bootstrapped AD test applied to the perihelion distribution finds, as expected,
that the uniform $q$ model can be strongly rejected at $>$99.99\% confidence. 

\section{Two-component Perihelion Distribution}\label{sec:splitq}

Evidently, the detached objects are \emph{not} uniformly distributed in
perihelion, even after accounting for bias. 
We observe only a small biasing
effect in the $q$ panel of \autoref{fig:unifqcumu}; there is a slight preference
towards detecting objects at low $q$, as expected, but we do not see any
indication of the sharp perihelion break that is observed in the real detached
population (\autoref{fig:aqdynclass}).

\begin{figure*}
    \centering
    \includegraphics[width=0.8\textwidth]{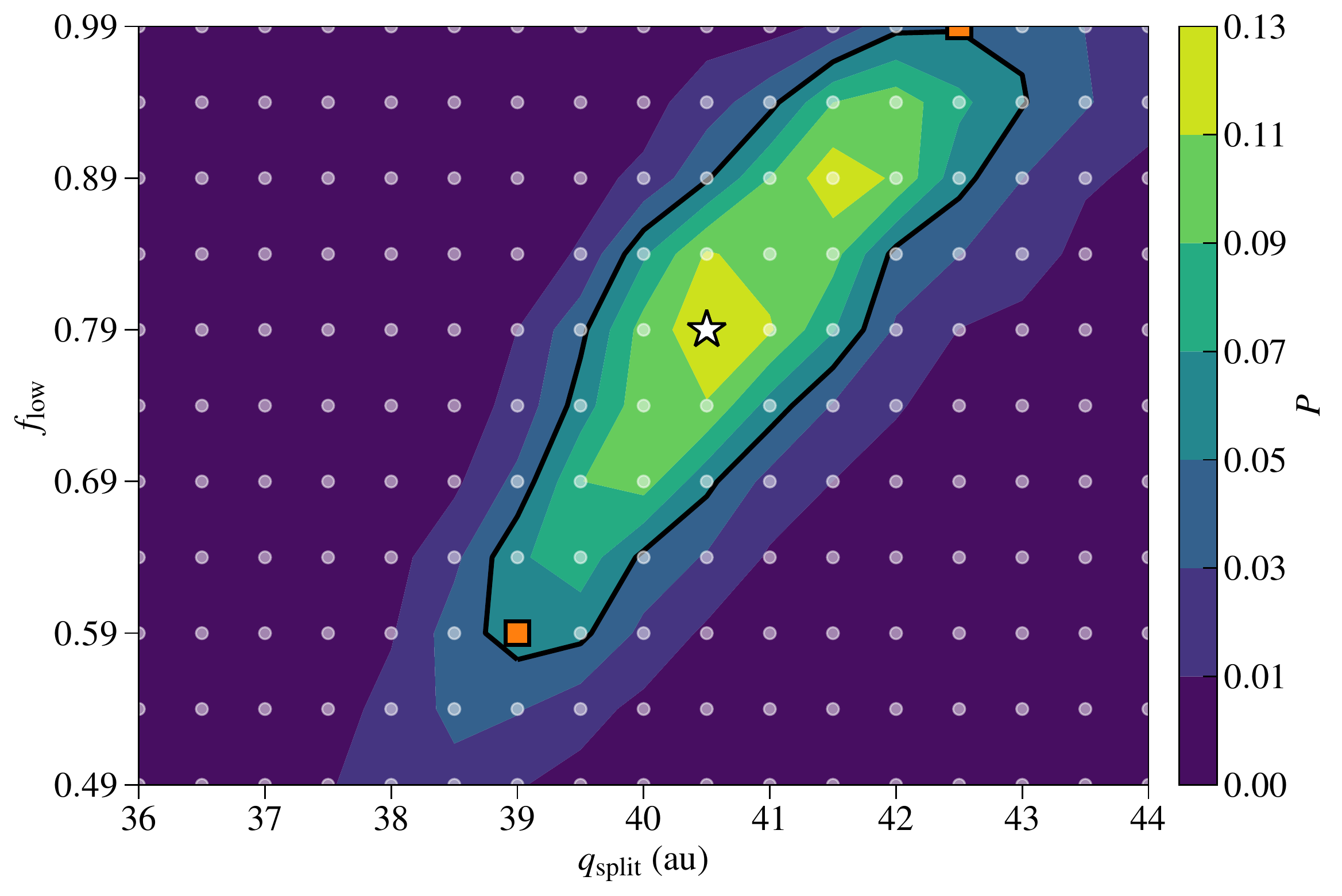}
    \caption{Contour plot of AD probability for a grid of $q_\mathrm{split}$ and
    $f_\mathrm{low}$ values in the two-component perihelion model.
    The solid black line indicates the boundary outside of which we can 
    reject a model at 95\% confidence or higher. The star represents the 
    $(q_\mathrm{split}, f_\mathrm{low})$ pair chosen for our least-rejectable
    distribution ($q_\mathrm{split} = 40.5$~au, $f_\mathrm{low} = 0.79$). 
    Orange squares indicate the two ``threshold'' cases shown in the middle and 
    bottom panels of \autoref{fig:threeq} and used for comparison
    population estimates in \autoref{sec:popest}. A uniform distribution (strongly
    rejected) would have $q_\mathrm{split} = 49$~au and $f_\mathrm{low} = 1.0$.}
    \label{fig:qgrid}
\end{figure*}

It is clear the data require a distribution that accounts for the
abrupt change in slope around $q \simeq 38.5$~au.
A simple increase in complexity is to introduce a two-component
distribution that is the union of two uniform distributions;
we use the label $q_\mathrm{split}$ as the
parameter that separates each component. 
That is, object perihelia
are distributed uniformly between $q_\mathrm{min}$ and $q_\mathrm{split}$
and uniformly between $q_\mathrm{split}$ and $q_\mathrm{max}$. 
The second parameter, $f_\mathrm{low}$, denotes the fraction of the
population in the ``lower'' uniform distribution between
$q_\mathrm{min}$ and $q_\mathrm{split}$. 
Note that $f_\mathrm{low} = 1$ corresponds to the single-component uniform case
with $q_\mathrm{max} = q_\mathrm{split}$.
Because the biased model must be capable of producing simulated detections with
$q$ as large as 49~au (the largest-$q$ OSSOS detached object),
we study a grid of models with a maximum value of $f_\mathrm{low} = 0.99$;
for $10^4$--$10^5$ simulated detections, such models provide sufficient 
detections in the high-$q$ tail to properly test the rejectability of the full $q$ range of 34--49~au.

\begin{figure*}
    \centering
    \includegraphics[width=0.85\textwidth]{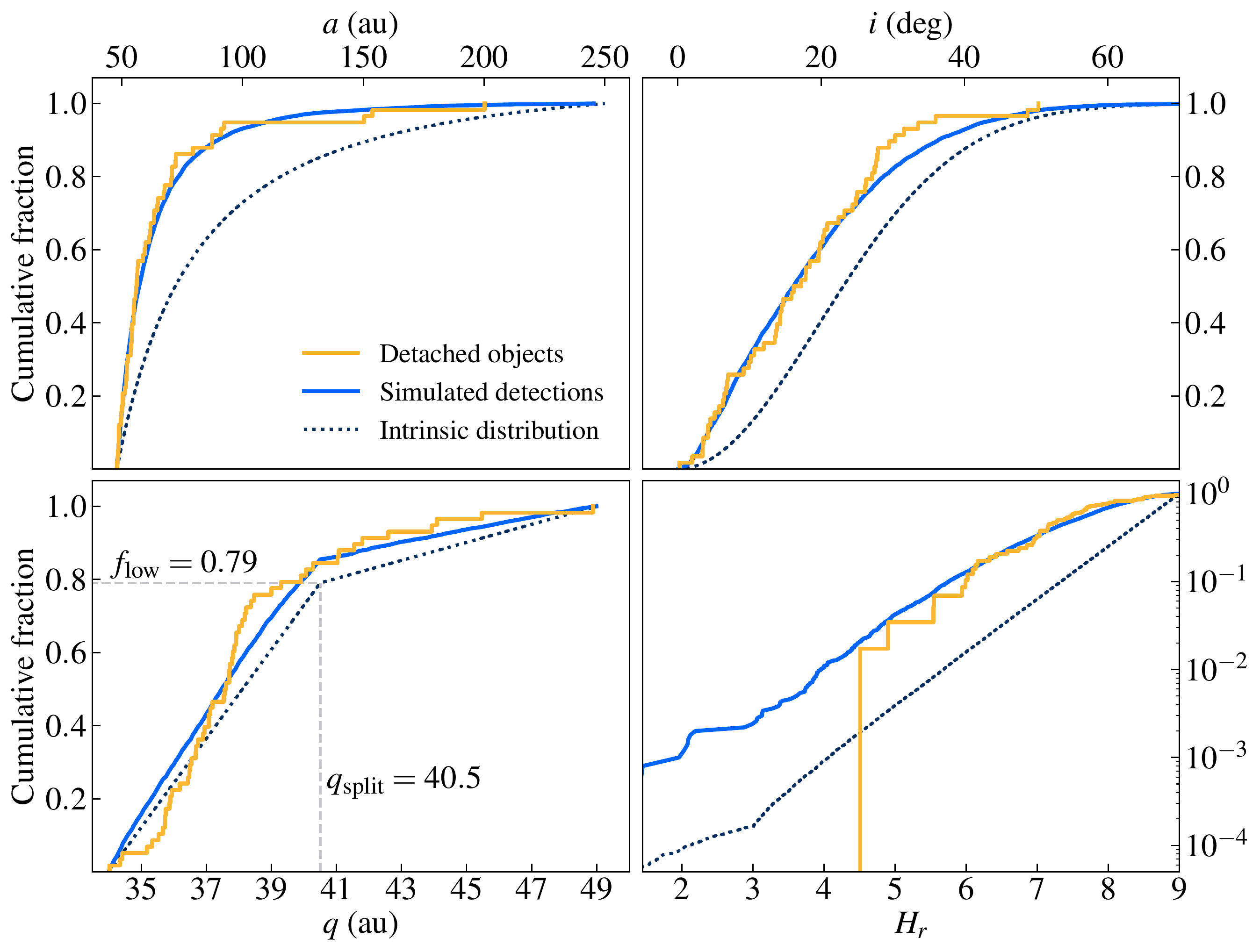}
    \caption{Cumulative distributions of the OSSOS detached objects (gold), simulated detections (blue), and the intrinsic model (dotted black curve)
    in the two-component $q$ model with the least-rejectable combination 
    $q_\mathrm{split} = 40.5$~au and $f_\mathrm{low} = 0.79$, 
    as indicated by the light-grey dashed lines in the lower-left panel. 
    AD probabilities for $a$, $q$, $i$, and $H$ and 89\%, 14\%, 65\%, and 
    26\%, respectively.
    We use this non-rejectable model to estimate the number of
    detached objects in \autoref{sec:popest}.}
    \label{fig:splitqcumu}
\end{figure*}

To determine the least-rejectable $(q_\mathrm{split}, f_\mathrm{low})$ pair, we
run a test similar to that of the inclination width except in two dimensions.
Over a grid with $q_\mathrm{split}$ ranging from 37.5--43~au in 0.5~au steps
and $f_\mathrm{low}$ ranging from 0.49--0.99 in steps of 0.05, we generate 10,000
simulated detections at each coordinate. We then use bootstrapped AD
statistics to evaluate the rejectability of the perihelion distribution
described
by each pair of parameters.
The results of this test are shown by the contour plot
of \autoref{fig:qgrid}. We are able to reject most combinations of
$(q_\mathrm{split}, f_\mathrm{low})$, with a correlation between them for the 
non-rejectable results. We choose the least-rejectable pair as our nominal
parametric perihelion distribution, which has
parameters $q_\mathrm{split} = 40.5$~au and $f_\mathrm{low} = 0.79$.
Importantly, no model described on this grid has AD test result above
14\%, which tells us in general that the ``uniform'' two-component $q$
distribution is an incomplete description of the detached objects.

Using the preferred parameters described above (and the prescribed $a$, $H$, and
$i$ distributions) we now generate 5,000 simulated detections to test the overall
acceptability of this model.
\autoref{fig:splitqcumu} shows the cumulative distributions of simulated detections
and the real objects for our preferred parametric model; none of the
distributions for $a$, $q$, $i$, or $H$ are rejectable in this case (see caption). 
Since we rejected the intrinsically uniform $q$ model but are unable to reject those with
slope breaks in the range $\sim$38--42~au, it must be the case that the
perihelion drop-off in the detached objects is a real physical phenomenon in 
the intrinsic population.

In the only other model used to estimate the observationally debiased number of detached
objects, \citet{Petit2011} implement a $q$ distribution borrowed from their
description of the hot main-belt classicals. It is mostly uniform between
35--40~au with a transition to a weak exponential tail beyond 40~au 
\citep[see Appendix A of][]{Petit2011}. Of course, this model was chosen with a small
sample of 13 detached objects. The known detached objects in our sample now
number to 58, but we still find that some sort of tapering beyond
$\sim$39--40~au in the intrinsic population is necessary to be consistent 
with current observations.

\begin{figure}
    \centering
    \includegraphics[width=\columnwidth]{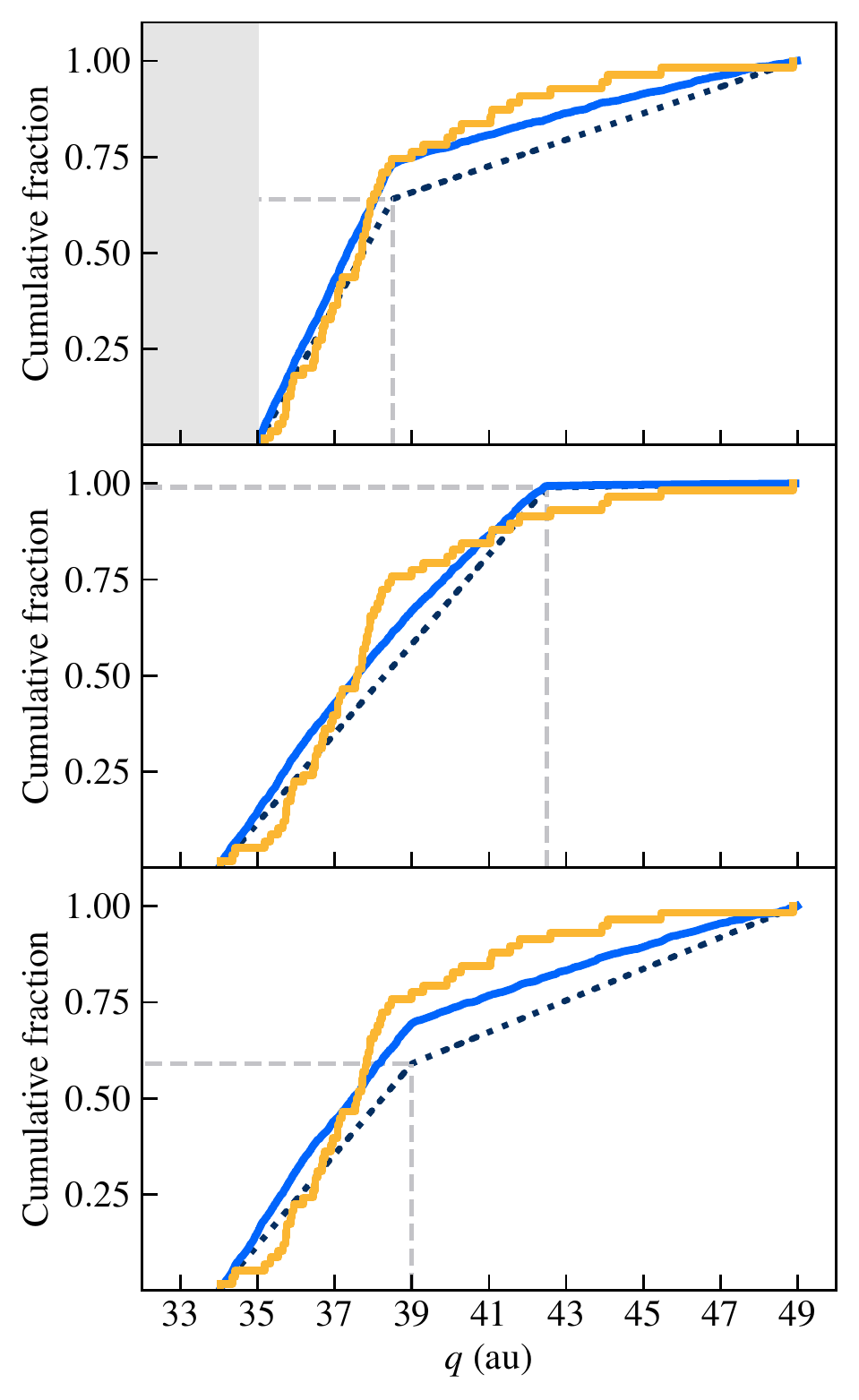}
    \caption{\textbf{Top panel:} a non-rejectable ($P = 27\%$) simulated 
    $q$ distribution plotted after having arbitrarily removed the $q < 35$~au
    detached objects (grey region). Without the extended low-$q$ tail seen in 
    the $q$ panel of \autoref{fig:splitqcumu}, the real detached distribution
    appears to be more similar to our heuristic two-component model with 
    $q_\mathrm{split} = 38.5$~au; 
    for $f_\mathrm{low} = 0.64$, one obtains a break in simulated detections 
    that is nearly identical to the one in the real detached distribution. 
    \textbf{Middle and bottom panel:} Two other perihelion distributions, both on the 
    threshold of rejectability ($P \simeq 5\%$), from the two-component model;
    the middle and bottom panels correspond to the upper right and lower
    left orange squares respectively in \autoref{fig:qgrid}. The inferior match
    of these models is illustrative of the quality of fit at the limit of acceptability.
    (Note that the $e$ distributions for these two cases are statistically acceptable ($P\approx50\%$) as 
    is the $a$ distribution, illustrating the superior diagnostic value of using the $q$ distribution as mentioned in 
    \autoref{sec:surveysim}).
    }
    \label{fig:threeq}
\end{figure}

We emphasize that we are not claiming this two-component uniform $q$ 
distribution is a wonderful description of the real detached objects. 
A more complex parametric model for $q$ could easily be developed. 
We simply use this as a relatively simple non-rejectable model to produce
the population estimates below, which is the main goal of this manuscript.
The reader might wonder why $q_\mathrm{split}=40.5$~au
has been selected by the tests, rather than something closer
to the 38.5~au value evident
in \autoref{fig:aqdynclass}.
The reason is simply that the small tail of $q<35$~au detached objects
forces the assumed linear behavior from the minimum of 34~au to be a 
poor representation of the low-$q$ portion of the distribution.
If we arbitrarily cut away the small $q < 35$~au section of the real and 
simulated samples, the same analysis now yields a cumulative $q$ 
distribution  (\autoref{fig:threeq}'s top panel)
with $q_\mathrm{split}=38.5$ that provides a better match to the 
distribution (with AD probability of 27\%) and looks visually more
like the real detections.

\subsection{Population Estimate}\label{sec:popest}
Our method for estimating the population (the debiased number of objects)
follows similar projects in the literature \citeeg{Kavelaars2009, Petit2011,
Gladman2012, Lawler2018, Ashton2021, Crompvoets2022}. 
For a given trial, we let
the survey simulator run until 
it detects the same number of detached objects as the real sample, 
recording how many objects it had to draw to do so; this drawn
number is the population estimate for that trial. 
We bin 1,000 trials and take the median to be our nominal population
estimate, and determine the 95\% confidence
interval from the 25th and 975th values in a sorted list (2.5\% wings on either
side of the distribution). This gives a result for population of detached
objects with $H_r < 8.66$ of 
\begin{equation}
N(H_r < 8.66) = \text{36,000}^{+12,000}_{-9,000}\quad\text{(parametric estimate)}
\label{eq:parametricpop}
\end{equation}
in the semimajor axis range 48--250~au. This is the
first observational population estimate of the detached objects based on
independently-determined orbit distributions and is an important goalpost for cosmogonic models.

To investigate the dependence of the population estimate on our specific
perihelion parameterization, we repeat the estimate procedure for two additional cases 
in the two-component perihelion model (orange squares in \autoref{fig:qgrid}, whose
$q$ distributions are shown the bottom two panels of \autoref{fig:threeq}). 
For $q_\mathrm{split} = 39.0$~au and
$f_\mathrm{low} = 0.59$, which according to \autoref{fig:qgrid}, is at the
threshold of rejectability (lower left end of the acceptable region); 
this model gives a median population of
$\text{37,000}^{+10,000}_{-9,000}$, within 3\% of the
least-rejectable model. 
Similarly, at the upper-right boundary, a model
with $q_\mathrm{split} = 42.5$~au and $f_\mathrm{low} = 0.99$ produces a very 
similar median population of
$\text{35,000}^{+11,000}_{-9,000}$.
This illustrates two important points. 
Firstly, it tells us that the population estimate is not strongly coupled to
the particular choice of $q$ distribution (despite the match obviously appearing
inferior in \autoref{fig:threeq}).
Secondly, it tells us that the
uncertainty on our population estimate is still dominated by the Poisson error
from the small number of known detached TNOs
rather than the specifics of a perihelion model.

\section{Survey Simulation of a Rogue Planet Model}\label{sec:rogue}
Having determined a non-rejectable parametric description of the detached
TNO orbit and size distributions,
we now switch our focus to another application of the same machinery.
We are able forward-bias orbital element models and produce a population
estimate without any parametric models. Using the previously established methodology,
we can compare the results of a numerical simulation to the OSSOS detached data set. As one
illustrative example, we consider the results of a numerical simulation \citep{Huang2022}
of a rogue planet scenario, which is a possible explanation for the perihelion
lifting of the detached objects \citep{Gladman2006}.
In such a model, an Earth-mass or greater planet forms
in the outer Solar System contemporaneously with the other planets; strong
giant planet interactions initially launch the rogue to large semimajor
axis followed by a period of weak perturbations while the rogue exists
on a meta-stable orbit in the scattering disk.
During this period, the rogue strongly influences the early TNOs, including the primordial scattering
population.
Such rogue planets are ejected via scattering on a 100~Myr time scale \citep{Gladman2006}.
Using the output of one particular rogue planet numerical integration, 
we compare the output of such a model to the observed orbital distributions 
of the detached objects discovered in OSSOS.

We have access to the results of a recent rogue planet simulation \citep{Huang2022}; 
it is important to note that the rogue's mass and orbit has not yet been 
tuned in any way to ``fit'' the observed detached objects.
The rogue is a two Earth-mass planet which starts the $q$-lifting simulation with
$a = 300$~au, $q = 40$~au, and $i = 20^\circ$. 
The Sun and giant planets start on their current orbits, mutually
interacting and affecting (weakly) the rogue's orbit.
There are $10^5$ test particles that represent the primordial 
scattering TNOs, which are initially distributed from 50--600~au 
following a $dn/da \propto a^{-2.5}$ power law. 
The initial $q_0$ of the test particles are uniformly assigned 
from 33--37~au, and the initial $i$ is distributed according to 
$\sin(i)$ times a Gaussian with width $\sigma_i = 15^\circ$ (as 
observed in the hot main belt classicals).

\begin{figure*}
    \centering
    \includegraphics[width=0.8\textwidth]{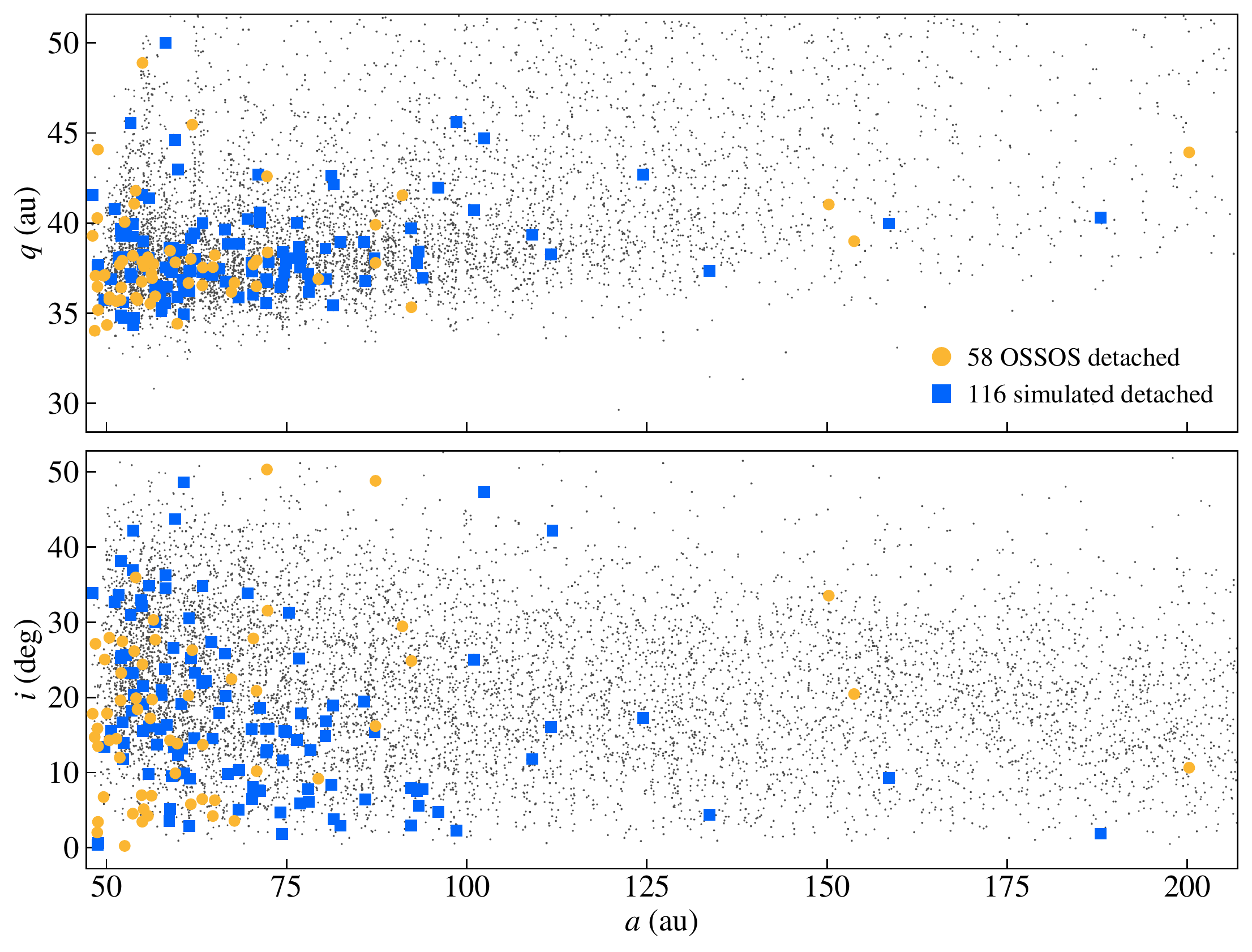}
    \caption{The 58 TNO OSSOS sample we are using, compared the intrinsic
    distribution (black dots) provided by the 4 Gyr end-state of a numerical
    simulation with a rogue planet that inhabited the distant solar system 
    for 100 Myr (see text). The blue squares show a set of (twice as many) 
    simulated detections from the intrinsic distribution, illustrating the 
    detection bias to lower $a$, $q$, and $i$.}
    \label{fig:roguecompare}
\end{figure*}

\begin{figure*}
    \centering
    \includegraphics[width=0.85\textwidth]{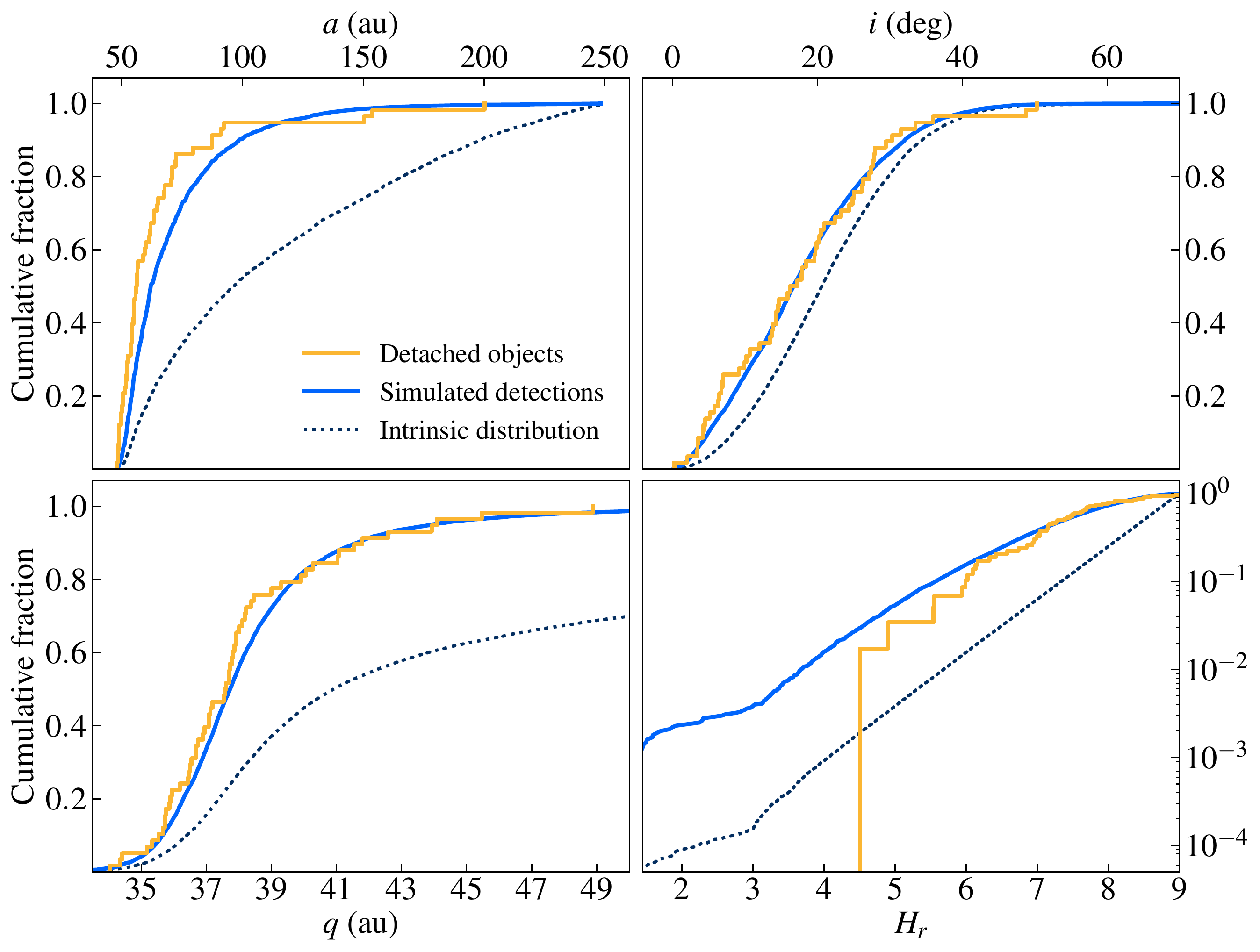}
    \caption{The intrinsic distribution and simulated detections coming from the end state of a numerical rogue planet model (see text).  This model has a larger fraction of objects at high-$a$ and high-$q$ orbits, but when the biased detections are compared to the real detached objects one sees that the perihelion distribution (lower left) provides an excellent match, with an AD probability of 40\%.} 
    \label{fig:roguecumu}
\end{figure*}

The initial conditions are then integrated forward 100~Myr using GLISSER, 
an improved GPU-based planetary system integrator based on \citep{Zhang2022}
but now with the capability to resolve test-particle close encounters
with the massive bodies.
Over the 100~Myr, the rogue's semimajor axis has variation of no more than
$\pm30$~au.
After 100~Myr, the rogue is manually removed from the system \citep[for more details, see][]{Huang2022}.
The system, sans rogue, is then integrated forward to 4~Gyr to bring
the orbital distributions to the present epoch for comparison with
observations; there are only mild modifications of the detached population because, 
after the rogue leaves, most $a<250$~au detached TNOs are 
``frozen into place''. 
The $q \lesssim 35$ scattering TNOs are especially heavily depleted as 
Neptune ``erodes'' this population, although a few can ``stick'' to 
resonances \citep{Duncan1997, Lykawka2007b, Yu2018}. 
Lastly, the surviving test particles are classified by the 
algorithm described in \citet{Gladman2008nomen}, which analyzes
the 10 Myr behavior of each particle. We use the ``detached'' test particles
with semimajor axes between 48~au and 250~au as the ``intrinsic model'' 
from which we draw objects during survey simulation.
\autoref{fig:roguecompare} shows the 4~Gyr end-state of the model, as well as the 58 
OSSOS TNOs and twice as many simulated detections (for illustrative purposes).
We do not here compare to objects classified as resonant because their
selection effects are more complicated. 
Although the simulated detached of \autoref{fig:roguecompare} clearly exhibit 
the bias to lower $a$, $q$, and $i$ as seen in the real detached TNOs, there 
appears to be comparatively more simulated detections at large $a$; this is shown
quantitatively in the cumulative distribution in \autoref{fig:roguecumu}.

Survey simulation of this rogue planet model was also performed by 
\citet{Huang2022}; however, we take a more rigorous approach and focus exclusively 
on the OSSOS detached sample at all $q$.\footnote{\citeauthor{Huang2022} used all 
$q > 38$~au objects available from the Minor Planet Center (MPC), but for 
this  sample precise debiasing is not possible.}
Selecting all the detached classifications with $48<a<250$~au from the model,
we used the survey simulator in lookup mode and applied the $H$ distribution
from \autoref{sec:aandHdists}. 
We generated 5,000 simulated detections based on
the detached object output of the rogue planet simulation. The cumulative 
distribution plot is shown in \autoref{fig:roguecumu}. 
The AD test results for $a$, $q$, $i$, and
$H$ are $0.01$\%, 40\%, 76\%, and 45\%, respectively.

We find that the action of the rogue planet generates $q$ and $i$ distributions 
that are non-rejectable with no significant exploration of the
rogue model's parameter space. 
Surprisingly, this first model is able to reproduce the whole detached perihelion
distribution (see the $q$ panel of \autoref{fig:roguecumu}) better than any of 
our parametric models.
Additionally, despite being initially distributed with a
colder inclination width of only $15^\circ$ (recall this was rejectable at
$>$99\% confidence in \autoref{sec:idist}), the rogue planet's effect provides
enough heating in $i$ to produce the large-$i$ tail and is an excellent match, 
despite not being tuned to do so.
The main mechanism by which $q$ and $i$ increase from the initial conditions is
Kozai cycling within mean-motion resonances \citep[reviewed by][]{Gomes2008}.
This secular effect causes a correlated increase in $q$ and $i$ when an object 
falls into an MMR of Neptune, and if the object drops out of the resonance
(either by Neptune migration or by changing $a$ due to close encounters with the rogue), 
the higher $q$ and $i$ are ``frozen'' in place.
Lastly, it is unsurprising that the $H$ distribution is non-rejectable, since
we are assigning $H$ values to model objects post-facto during survey simulation. 

There is, however, a significant discrepancy between the real detached semimajor axis 
distribution and the survey simulated model distribution (which is strongly 
rejected). 
The intrinsic distribution from the simulated model contains a smaller fraction of objects 
at low $a$, which results in fewer simulated detections at low $a$ when compared 
to the real detached (see \autoref{fig:roguecumu}). 
The discrepancy at the small-$a$ tail of the distribution is the cause 
of the extremely small AD test probability (0.01\%), as the AD test is 
very sensitive to the tails. 
For comparison, the KS test (less sensitive to the tails) for $a$ indicates
order of magnitude less rejectability than the AD result, 
suggesting that the small-$a$ tail is indeed the issue even though the
KS test indicates a probability $<$1\%.
We find that the low-$a$ discrepancy is entirely due to the region between
the 2:1 and 5:2 MMRs;
if we restrict to only real and simulated detections with $a > 57$~au, then 
the $a$ distribution problem disappears (with AD probability 57\%), and none
of $q$, $i$, or $H$ are made rejectable either.
This mismatch indicates this sample numerical simulation
is missing some physics (although it is possible some
stochastic evolution in the rogue's history could increase
implantation $a<57$~au TNOs.
More likely, the simulation used as an example here lacks the
physics of migration, which has been shown
\citep{Nesvorny2016} to be especially effective at
populating semimajor axis range, due to the abundance
of mean-motion resonances.

Based on this cosmogonic model (in which the orbital elements distributions are
thus not fit to each orbital parameter), the usual population analysis yields
an estimated detached TNO population (with $H_r < 8.66$ and $48 < a < 250$~au) of $\text{48,000}^{+15,000}_{-12,000}$.
This estimate is larger than \autoref{eq:parametricpop} because the rogue model 
possesses a set of larger-$q$ TNOs that are \emph{extremely} difficult to detect;
the existence of such TNOs would require more intrinsic objects for the survey simulator to detect the 58 real TNOs.
\autoref{fig:roguecumu} shows that this rogue model has roughly one third of the 
intrinsic TNOs with $q > 50$~au and these are not expected to be detected in 
a survey even as large as OSSOS.
We believe it very likely that in reality there are $q>50$~au TNOs in 
the $a<250$~au region, as predicted by numerical simulations
\citep{Gladman2006, Lykawka2008, Nesvorny2016, Huang2022}, which thus forces up
our population estimate. 

\section{Discussion and Conclusions}\label{sec:discussion}

We first discuss similarities of our detached population estimate to the few
that exist in the literature, and then compare it to populations of other dynamical 
classes, for context.

\subsection{Comparison to other detached estimates}
\label{sec:comparedet}

Our parametric detached population estimate
(\autoref{eq:parametricpop})
was 36,000 TNOs, with a variation due to the
model systematics which is smaller than the 30\% variation coming
from the Poisson statistics.
The orbital distribution coming from 
the single rogue planet simulation we examined
had a portion of the detached population with harder-to-detect
orbits, but even this only increased to the estimate to
48,000, again with 30\% Poisson uncertainties (at 95\% 
confidence).

Because of these variations, and because some $q>50$ au component is 
extremely likely to exist, we believe our results justify only a single 
significant figure and we estimate to roughly 50\% accuracy:
\begin{equation}
N(H_r < 8.66) = (5\pm2) \times 10^4 
\quad
(48 < a < 250~\text{au}) .
\label{eq:finalpop}
\end{equation}
The only systematic effect that could likely invalidate this is if there is 
an enormous
hidden population of $q \gg 50$~au detached objects with $a<250$~au.
Observations indicate a lack of $q>50$~au objects
below some $a$ threshold, which has been used as 
a constraint on the production mechanism of ``Sednoids''
\citep{Morbidelli2004, Gladman2006, Trujillo2014}.
The recent Dark Energy Survey \citep{Bernardinelli2022} found 
no $q>50$~au detached objects\footnote{2014 US$_{277}$ is 
likely
to be in the 6:1 resonance (P.~Bernardinelli, private communication, 2022).}
out to $a=500$~au.

The only other direct observational estimate of the full $48<a<250$~au detached population is from the CFEPS survey; 
\citet{Petit2011} estimated $N(H_g < 8) =
\text{10,000}^{+7,000}_{-5,000}$ for all non-scattering TNOs with $a > 48$~au 
(assuming a $dn/da \propto a^{-2.5}$ distribution, which is thus
effectively an estimate for $a<250$~au since only 8\% of the TNOs
are beyond 250~au for that semimajor axis).
We adjust this to $H_g$ to $H_r$ by adopting their assumed $g - r = 0.7$ color, 
making their 10,000 estimate correspond to $H_r < 7.3$. 
CFEPS extrapolated down to 100~km diameter ($H_r\simeq 8.66$) by using 
the same (steep) $\alpha=0.8$ slope as for their relatively bright detections, but
deeper surveys prefer shallower values of $\alpha$ 
\citep{Fraser2014, Kavelaars2021}.
For consistency, we scale from $H_r=7.3$ to 8.66 using our $\alpha=0.6$ and keep
92\% of the population to restrict to $a<250$~au, resulting
in the \citeauthor{Petit2011} study suggesting 60,000 detached TNOs, with a factor
of 2 error bar just from the statistics and some additional uncertainty
from the unknown $\alpha$.
This estimate is thus consistent with our new study which, unlike
\citeauthor{Petit2011}, has been matched to the better-determined
orbital distribution now that there are more known detached TNOs.

Although not based directly on observations of detached objects,
\citet{Nesvorny2016} used the results of their outer Solar System emplacement
model caused by grainy outward Neptune migration to estimate a detached
population of 40,000 $D>100$~km TNOs with a more restricted semimajor axis range
of $50<a<100$~au.
Since we have concluded the detached $a$ distribution follows $dn/da\propto a^{-2.5}$,
this trims our 50,000 $48<a<250$~au distribution to 33,000 and is thus very 
consistent with the \citeauthor{Nesvorny2016} estimate.
Although this grainy migration model was not compared to the details of the 
orbital distribution of detached objects, the number match is impressive given that
the absolute normalization was set by the need to capture enough Jupiter trojans
in the same process.
A comparison incorporating bias needs to be done for  the $a$, $q$, and $i$ distributions
of these models. However, there also appears to be an obvious problem when looking at 
the detached to resonant ratio (see below).

\subsection{Comparison to other dynamical classes}
\label{sec:comparedyn}

A proposed cosmogonic model should be able to both produce the  absolute population  of the detached component, as well as  its relative numbers compared to the hot main  belt,  
and the resonant and  scattering components in the same semimajor axis range, 

\paragraph{Hot main Kuiper belt.}
The hot main belt estimate from \citet{Petit2011} is 
$N(H_r < 7.3) = \text{4,100}^{+900}_{-800}$; scaling by a factor of $10^{0.6(8.66 - 7.3)}$ 
gives  $N(H_r < 8.66) = \text{27,000}^{+6,000}_{-5,000}$.
We thus estimate that detached objects outnumber the main belt by a factor of 
$\simeq$2,  comparable to \citet{Petit2011}'s ratio of 2.5. 
If the detachment of ancient scattering TNOs into the hot main belt is 
comparably efficient to that beyond the 2:1, one might expect that the
ratio of $a>48$~au detached to hot main belt would just follow the 
ancient semimajor axis distribution.
For an $a^{-2.5}$ power law, the ratio of numbers from $48<a<250$~au 
to $40<a<48$~au is 3.
This is possibly interesting as a constraint on the detachment physics,
and continues to be in line with the ``unification'' model of 
\citeauthor{Petit2011} (their Sec.~5.2.3) that supports all the hot
populations derive from a single cosmogonic process.

\paragraph{Today's scattering population.}
\citet{Petit2011} used the CFEPS survey to estimate $\simeq$5,000 $D>100$~km 
scattering TNOs, with a factor of at least 3 error bar.
The OSSOS-based estimate for scattering TNOs \citep{Lawler2018} is an order
of magnitude larger, at 80,000, but covers the entire scattering population
out to very large $a$; we retrieved that model and determined that $48<a<250$~au
represented $\simeq$40\% of the model, and thus in this semimajor axis range
the detached TNOs outnumber the scattering by a ratio of only
$5\times10^4/\text{32,000}\simeq1.5$.
We are surprised the ratio is this small.
The rogue model suggests an intrinsic detached/scattering ratio 
(today, after dynamical erosion of the  age of the solar system) 
of 10:1, which would match the ratio based on the 
\citeauthor{Petit2011} scattering estimate.
For comparison, \cite{Nesvorny2016} estimate this ratio should be 4--5.
Revisiting these studies is clearly appropriate now that a reliable detached
population estimate exists to compare to, and should be an additional
constraint on the detachment process(es).
    
\paragraph{The nearby resonant populations.}
An obvious dramatic disagreement seems to be contained in the realization
\citep{Gladman2012}
that the distant resonant populations are huge, with many thousands of 
objects in each of the main distant resonances.
Most recently, \citet{Crompvoets2022} used OSSOS to estimate the distant 
resonances (beyond the 2:1) contain at least 100,000 objects with $H_r < 8.66$, 
albeit with a factor of four uncertainty (at 95\% confidence).
Thus, in the $48<a<250$~au range, the detached/resonant ratio is $<$0.5; to
our knowledge no emplacement models to date generate enough resonant objects
at large $a$.
The rogue model tested here has a ratio of 2.3 (that is, fewer resonant 
than detached).
\citet{Nesvorny2016}'s two grainy models give 8 or 11 for the 
ratio of $50<a<100$~au detached to the total of the four $n$:1 resonances
in that range;
our detached estimate in that range is 33,000 TNOs, while the sum from
\citeauthor{Crompvoets2022} of the 3:1, 4:1, 5:1, and assuming 10,000 in the 6:1
resonance is 51,000 gives a detached/resonant ratio of $\simeq$0.65.
Thus, the huge current TNO population trapped in large-$a$ resonances
is an outstanding mystery and we are not aware of any models to create
detached TNOs that simultaneously produce sufficiently large resonant
populations in the same semimajor axis range.

\subsection{Conclusions}
The fact that the current detached population is nicely described by the
$dn/da \propto a^{-2.5}$ power law generates some interesting thought
experiments.
The theoretical long-term steady-state for a population of scattering
objects that starts with a single semimajor axis
\citep{Yabushita1980, Duncan1987, Malyshkin1999}
has a projected surface density $\Sigma(a) \propto a^{-2.5}$,
corresponding to a shallower $dn/da \propto a^{-1.5}$; 
this is the distribution for the scattering disk after 4 Gyr of
evolution in numerical simulations
\citep{Levison1997}.
Several possible scenarios are possible.
We have confirmed numerically that $\sim$30~Myr after
Neptune begins scattering TNOs on nearby orbits out beyond
$a=50$~au, the power law is still steeper than the steady state 
and we find $dn/da \propto a^{-2.5}$ at this time (hereafter
the power laws are all for $dn/da$); if $q$ lifting
occurs across the entire $48<a<250$~au range with a semimajor axis
independent efficiency, then -2.5 could be preserved.
Alternately, if the steady state ($-1.5$) had been reached, then
if the efficiency of $q$ lifting went as $1/a$ the $-2.5$ index 
could be reached.
Lastly, if the perihelion lifting occurred extremely early when
the index as even steeper than $-2.5$ the probability of lifting
$q$ would have to increase with $a$ (this would likely be the
case for secular perihelion oscillations as $a$ approaches that
of the rogue).
All of these scenarios are likely oversimplifications because
resonance trapping \citep{Gomes2008, Nesvorny2016, Huang2022}
certainly plays a role.
Only careful comparisons between the final (4.5~Gyr) states of 
cosmogonic numerical simulations and calibrated observations
surveys with bias measurements (like the preliminary comparison
we did here between a rogue-planet model and OSSOS) are likely
to be convincing as observational surveys advance and numerical
studies improve.

The population comparisons to other dynamical classes (\autoref{sec:comparedyn})
are made possible by now having (what we believe
to be) a reliable estimate of the detached population's numbers, albeit
still with 50\% uncertainty at 95\% confidence.
Since the systematics seem to be under control (with model variations
causing population variations less than the 95\% confidence intervals)
more characterized detections will be required to improve the estimate.
The LSST project should certainly provide better statistics of detached
detections.

With our number estimate for the detached objects, a (more uncertain) 
total mass estimate can be made (in this $a$ range and to a limiting  
magnitude/size).
The mass will be dominated by the large number of objects with 
$3 < H_r < 8.66$, which, if we adopt a 17\% albedo for detached TNOs 
\citep{Santos-Sanz2012}, corresponds to radii of 770~km and 57~km.
Fewer than 0.02\% of the objects have $H_r < 3$  ($D > 770$~km) for
our assumed $H$-magnitude distribution.
Writing the differential size distribution $dn/dR = kR^{-(5\alpha_* + 1)}$ 
(see footnote in \autoref{sec:aandHdists}) with $\alpha_* = 0.6$, we obtain
the normalization factor $k$ by integrating this distribution to give our 
population of $5\times10^4$.
Integrating the differential mass distribution yields 
\mbox{$M \sim \rho_\mathrm{cgs} \cdot 3\times10^{26}~\text{cm}^3$}, or 
$M \simeq 0.05~M_\oplus$ using $\rho_\mathrm{cgs} = 1.0$~g/cm$^3$, with a
factor of at least 3 uncertainty due to assumptions on the albedo and density.  
This just says that the detached
population's total mass is comparable to the main belt.

We are impressed by the untuned improvement in the perihelion
distribution that resulted from using the rogue planet simulation's
output at the present epoch.
Caution needs to be taken about the quality of the survey characterization
in future studies; the details of the shape of the detected perihelion distribution
of detached TNOs may be one of the strongest constraints on the detachment
mechanism, and understanding the survey detection biases with high precision
may be important to discriminate between models.
Luckily we are entering the era where the number ratio and detailed orbital
distributions for the $a>48$~au dynamical populations are becoming measurable
with better than factor of two precision.
Improved observational estimates of the resonant and scattering should also
be forthcoming, and are arguably more impactful in the short term.

\section{Acknowledgements}
We acknowledge Canadian funding support from NSERC (grant 2018-04895).
YH acknowledges support from China Scholarship Council (grant 201906210046) and the Edwin S.H. Leong International Leadership Fund.

\bibliography{manuscript}{}
\bibliographystyle{aasjournal}



\end{document}